\documentclass[aps,pra,onecolumn,amsfonts,amssymb,a4paper,showpacs]{revtex4}

\usepackage{graphicx}

\newcommand{\be}{\begin{equation}}
\newcommand{\ee}{\end{equation}}

\newcommand{\ket}[1]{|{#1}\rangle}
\newcommand{\braket}[2]{\langle{#1}|{#2}\rangle}

\newcommand{\nbar}{{\bar n}}

\newcommand{\sumlim}{\sum\limits}

\newcommand{\ehoch}[1]{e^{#1}}

\newcommand{\drm}{{\text{d}}}

\newcommand{\skipc}[2]{}

\begin{document}

\title{Wave packet dynamics and factorization of numbers}

\author{Holger~Mack}
\author{Marc~Bienert}
\author{Florian~Haug}
\author{Frank S.~Straub}
\author{Matthias~Freyberger\skipc}
\author{Wolfgang P.~Schleich}

\affiliation{Abteilung f\"ur Quantenphysik, Universit\"at Ulm, 89069 Ulm, Germany}


\begin{abstract}
We connect three phenomena of wave packet dynamics: Talbot images, revivals of a particle in a box and fractional revivals. The physical origin of these effects is deeply rooted in phase factors which are quadratic in the quantum number. We show that the characteristic structures in the time evolution of these systems allow us to factorize large integers.
\end{abstract}

\maketitle

\section{Talbot effect, revivals and factorization}
\begin{quote}
\textit{
A{\scriptsize LTHOUGH} so much has been explained in optical science by the aid of the undulatory hypothesis, yet when any \textsl{well-marked ph\ae{}nomena} occur which present unexpected peculiarities, it may be of importance to describe them, for the sake of comparison with the theory.
}
\end{quote}

This quotation is the opening sentence of a paper \cite{bib:talbot} by H.F.~Talbot entitled {\it Facts relating to optical sciences}, published in 1836. In this article Talbot reports his experiments on interference of light. We quote again from his seminal paper:

\begin{quote}
\textit{
About ten or twenty feet from the radiant point, I placed in the path of the ray an equidistant grating$^*$ made by Fraunhofer, with its lines vertical. I then viewed the light which had passed through this grating with a lens of considerable magnifying power. The appearance was very curious, being a regular alternation of numerous lines or bands of red and green colour, having their direction parallel to the lines of the grating. On removing the lens a little further from the grating, the bands gradually changed their colours, and became alternately blue and yellow. When the lens was a little more removed, the bands again became red and green. And this change continued to take place for an indefinite number of times, as the distance between the lens and grating increased. In all cases the bands exhibited two complementary colours.
It was very curious to observe that though the grating was greatly out of the focus of the lens, yet the appearance of the bands was perfectly distinct and well defined.}\\

\textit{
\footnotesize
$^*$A plate of glass covered with gold-leaf, on which several hundred parallel lines are cut, in order to transmit the light at equal intervals.
}
\end{quote}

In today's language Talbot has considered the diffraction of light from a grating in the near-field zone. He found that the intensity distribution immediately after the grating repeats itself periodically at multiples of a characteristic distance, later called Talbot length. Moreover, at fractions of the Talbot length the pattern reappears in a rescaled version. Indeed, at the fraction $1/r$ of the Talbot length we find $r$ substructures of the original pattern. The theoretical explanation of this phenomenon was provided in 1881 by Lord Rayleigh \cite{bib:rayleigh}.

The Talbot effect occurs not only for light waves, but also for matter waves. It has been observed for atoms \cite{bib:talbotexperiments}, and most recently for $\text{C}_{60}$ molecules \cite{bib:c60talbotexperiment}. For a comprehensive review of the Talbot effect, we refer to ref.~\cite{bib:rohwedder}.

A phenomenon closely related to the Talbot effect occurs in the time evolution of wave packets \cite{bib:buch}, such as a Rydberg electron. An initially well-localized packet spreads over its orbit but regains its original shape after a characteristic time, which is much larger than the classical period. Moreover, at fractions of this revival time, the wave packet splits into multiple copies. This effect of fractional revivals appears most clearly in the well-known problem of the particle in a box \cite{bib:stifter, bib:friesch}. Revivals and oscillations can also be used for a new type of interferometer for light in a planar multimode waveguide \cite{bib:revivals}.

Fractional Talbot images as well as the fractional revivals originate from phase factors which are quadratic in the quantum number. They give rise to Gauss sums \cite{bib:lang}, which have interesting number-theoretical properties.

This immediately points to the newly emerging field of quantum information processing \cite{bib:books}, which has received an enormous drive from the discovery of the Shor algorithm \cite{bib:shor1,bib:shor2} to factorize large numbers --- a paradigm of number theory. The existence of distinct phenomena at fractions of a characteristic time suggests that the Talbot effect or wave packet physics may have links to the problem of factorizing numbers. Indeed, a recent proposal \cite{bib:dowling} makes use of a $N$-slit Young interferometer, where $N$ is the number to be factorized. Such a device is described by the Talbot effect due to a finite grating. The interference structure along the transverse direction at a fixed position after the grating serves as an indicator for a factor. For another approach we refer to ref.~\cite{bib:harter}.

In the present notes we show that the quadratic phase factors inherent in the time evolution of many quantum systems provide a tool to find the prime factors of a large number. Our article is organized as follows: In section~\ref{sec:talbot} we briefly summarize the essential ingredients of the Talbot effect in the language of atom optics. We devote section~\ref{sec:mathoftalbot} to the calculation of the free propagator of an array of wave packets. In section~\ref{sec:particleinbox} we connect this Green's function with the one of the particle in a box. In both cases we arrive at quadratic phase factors. We then show in section~\ref{sec:wavepackets} that similar phases appear in many discrete quantum systems. They manifest themselves in the autocorrelation function. We dedicate section~\ref{sec:fracrevivals} to cast the relevant sum into a form which brings out most clearly the phenomenon of fractional revivals. This form allows us in section~\ref{sec:factorization} to test our factorization scheme. We suggest a different approach towards factoring in section~\ref{sec:gausssums} based on properties of Gauss sums. We conclude in section~\ref{sec:conclusions} by presenting a brief outlook.

\section{Model of Talbot effect\label{sec:talbot}}
In order to set the stage for the mathematical treatment of the Talbot effect presented in the next section, we now define the principle setup. We consider the diffraction of an atomic wave from a grating. In principle our treatment is also correct for electromagnetic waves in paraxial approximation. Indeed, in this limit the d'Alembert wave equation reduces to the Schr\"odinger equation. For the influence of the higher order corrections we refer to ref.~\cite{bib:berryklein}. 

The grating, aligned along the $x$-axis, could be a mechanical or an optical one. Since we are interested in the subsequent propagation, it suffices to assume that the grating creates a wave function $\phi(x)$ with the period $d$ of the grating. In the case of an infinite array of slits, whose width is much smaller than their separation, the wave function
\begin{equation}
	\phi(x)=\sum_{n=-\infty}^{\infty}\varphi(x-n d)
\end{equation}
consists of an infinite number of independent initial wave packets $\varphi(x)$ separated by $d$. 

We assume that initially the atomic wave is under normal incidence, that is the wave vector $\vec k$, aligned along the $z$-azis, is orthogonal to the grating. When we consider an atom with an energy much larger than the recoil energy of the grating we can treat its motion along the $z$-axis classically. In this case time translates into the $z$-coordinate of the atom \cite{bib:yakovlev}, that is $z=v t$ where $v$ is its velocity along the $z$-direction.

We now consider the distribution of atoms along the $x$-axis for fixed propagation time $t$. For a defined velocity $v$ of the incoming atoms this time corresponds to a fixed position $z$ behind the grating. At the Talbot time $T$, that is at the distance $z_T\equiv v T$ the initial wave function $\phi(x)$ repeats itself. At time $T/2$ the interference pattern is identical to the initial one but shifted by half a period. At fractions of the Talbot time the period of the initial wave packet is a fraction of the original one. In the following sections we derive these results and show that they are a consequence of an intricate interference of phases.

\section{Mathematics of the Talbot effect}
\label{sec:mathoftalbot}

In the present section we briefly review the mathematical treatment of the Talbot effect. Our key tool is the Poisson summation formula. This analysis provides the foundation for the next section, where we emphasize the close connection between the Talbot effect and the particle in a box.

\subsection{Free time evolution of a periodic structure}

We consider the free motion of the wave packet
\begin{equation}
	\phi(x) = \sum_{n=-\infty}^{\infty}\varphi(x-n d)
	\label{eq:star}
\end{equation}
consisting of an array of infinitely many identical partial waves $\varphi(x)$ separated by a distance $d$. The propagator \cite{bib:buch}
\begin{equation}
	G_{\text{free}}(x,t|y,t=0)\equiv{\mathcal N}(t)\,e^{i\alpha(t)(x-y)^2}
	\label{eq:2star}
\end{equation}
of the free particle of mass $M$ with the scaling factor
\begin{equation}
	\alpha(t) \equiv \frac{M}{2\hbar t}
	\label{eq:alpha}
\end{equation}
and the normalization
\begin{equation}
	{\mathcal N}(t) \equiv \sqrt{\frac{M}{2\pi i \hbar t}} = \sqrt{\frac{\alpha(t)}{\pi i}}
	\label{eq:normalizationconst}
\end{equation}
allows us to find the wave function
\begin{equation}
	\psi(x,t)=\int\limits_{-\infty}^\infty\drm y\,G_{\text{free}}(x,t|y,t=0) \phi(y)
	\label{eq:3star}
\end{equation}
at a later time $t$.

Indeed, when we substitute the initial wave function, eq.~(\ref{eq:star}), and the Green's function, eq.~(\ref{eq:2star}), into the propagation equation, eq.~(\ref{eq:3star}), we arrive at
\begin{equation}
	\psi(x,t) = \int\limits_{-\infty}^{\infty}\drm y\,{\mathcal N}e^{i\alpha(x-y)^2} \sum\limits_{n=-\infty}^\infty\varphi(y-n d) = \int\limits_{-\infty}^{\infty}\drm\bar y\,{\mathcal N}\sum\limits_{n=-\infty}^\infty\,e^{i\alpha(x-\bar y - n d)^2}\varphi(\bar y)
\label{eq:starex}
\end{equation}
where in the last step we have introduced the integration variable $\bar y\equiv y-nd$.

Therefore, eq.~(\ref{eq:starex}) takes the form
\begin{equation}
	\psi(x,t)=\int\limits_{-\infty}^{\infty}\drm y\,G_{\text{Talbot}}(x,t|y,t=0) \varphi(y)
\end{equation}
where
\begin{equation}
	G_{\text{Talbot}}(x,t|y,t=0)\equiv{\mathcal N}(t) \sum\limits_{n=-\infty}^{\infty} e^{i \alpha(t) (x-y-n d)^2}
	\label{eq:propgt}
\end{equation}
denotes the Talbot propagator.

\subsection{Quadratic phase factors}

In order to bring out the relation of $G_{\text{Talbot}}$ to the propagator $G_{\text{box}}$ of the particle in a box, discussed in the next section, we rewrite eq.~(\ref{eq:propgt}) with the help of the Poisson summation formula
\begin{equation}
	\sum\limits_{n=-\infty}^{\infty} f_n = \sum\limits_{m=-\infty}^{\infty}\int\limits_{-\infty}^{\infty}\drm n f(n) e^{-2 \pi i m n}
\end{equation}
where $f(n)$ is a continuous extension of $f_n$ such that $f(n)=f_n$ at integer values $n$. 

This relation allows us to replace the summation by a sum of Fourier integrals. Hence, the Green's function $G_{\text{Talbot}}$, eq.~(\ref{eq:propgt}), takes the form
\begin{eqnarray}
	G_{\text{Talbot}}(x,t|y,t=0)&=&\sum\limits_{m=-\infty}^{\infty}{\mathcal N}\int\limits_{-\infty}^{\infty}\drm n\,e^{i\alpha (x-y-n d)^2} e^{-2\pi i m n}\nonumber\\
	&=&\sum\limits_{m=-\infty}^{\infty}\frac{1}{d}\,e^{-i \kappa_m(x-y)}\sqrt{\frac{\alpha}{i\pi}}\int\limits_{-\infty}^{\infty}\drm\xi\,e^{i\alpha \xi^2} e^{i\kappa_m \xi},
\end{eqnarray}
where we have introduced the new integration variable $\xi \equiv x-y-nd$ and the wave number
\begin{equation}
	\kappa_m \equiv m \frac{2\pi}{d}.
\end{equation}
Moreover, we have recalled the definition, eq.~(\ref{eq:normalizationconst}), of the normalization ${\mathcal N}(t)$.

When we perform the Gauss integral
\begin{equation}
	\sqrt{\frac{\alpha}{i\pi}}\int\limits_{-\infty}^{\infty}\drm\xi\,e^{i\alpha\xi^2} e^{i\kappa_m\xi}=\exp\left[-i\frac{\kappa_m^2}{4\alpha}\right],
\end{equation}
and make use of the definition, eq.~(\ref{eq:alpha}), of the parameter $\alpha$, we can identify the phase
\be
	\frac{\kappa_m^2}{4\alpha}=\frac{(\hbar \kappa_m)^2}{2M}\frac t\hbar\equiv E_m\frac t\hbar\equiv 2\pi m^2\frac tT.
\ee
Here we have introduced the energies 
\be
	E_m \equiv \frac{(\hbar \kappa_m)^2}{2M}
\ee	
and the Talbot time
\be
	T \equiv \frac{Md^2}{\hbar\pi}.
\ee

Hence, the propagator $G_{\text{Talbot}}$ takes the form
\be
	G_{\text{Talbot}}(x,t|y,t=0)=\frac1d\sumlim_{m=-\infty}^{\infty}\exp\left[-im\frac{2\pi}d(x-y)\right]\exp\left(-2\pi im^2\frac tT\right).
	\label{eq:gtalbot}
\ee
In the time dependent phase factor we recognize the quadratic dependence on the summation index $m$. 

\subsection{Integers and half integers of Talbot time}
According to eq.~(\ref{eq:gtalbot}) the propagator for $t=s\cdot T$ with an integer $s$ is identical to the propagator at time $t=0$. Consequently, at integer multiples of the Talbot time the wave packet regains its initial shape.

Furthermore, at $t=(s+\frac12)\cdot T$ the relation
\be
	\exp\left(-2\pi im^2(s+\frac12)\right)=\exp\left(-\pi im^2\right)=(-1)^m=\exp\left(-im\pi\right)
\ee
casts the propagator into the form
\be
	G_{\text{Talbot}}(x,t=(s+1/2)T|y,t=0)=\frac1d\sumlim_{m=-\infty}^{\infty}\exp\left[-im\frac{2\pi}d\left(x-y+\frac d2\right)\right].
\ee
We can identify the right-hand side of this equation with the propagator at time $t=0$ shifted by half the period $d$. Hence, the initial wave packet repeats itself at half integer multiples of the Talbot time but is displaced by $d/2$.

\section{Particle in a box}
\label{sec:particleinbox}

We now turn to the standard problem of a particle of mass $M$ in a box of length $L$. We first briefly review the essential equations and then derive the corresponding propagator. We conclude by comparing this result to the Talbot propagator.

\subsection{Propagation}

The time evolution of an initial wave packet $\psi(x,t=0)\equiv\varphi(x)$ in a box reads
\be
	\psi(x,t)=\sumlim_{n=1}^\infty\psi_nu_n(x)\ehoch{-iE_nt/\hbar}
	\label{eq:psiofxandt},
\ee
where the expansion coefficients
\be
	\psi_n\equiv\int\limits_0^L\drm y\,\varphi(y)u_n(y)
	\label{eq:kringel}
\ee
are in terms of the energy eigenfunctions
\be
	u_n(x)\equiv\sqrt{\frac2L}\sin(k_nx)
\ee
with wave numbers
\be
	k_n\equiv n\frac\pi L
	\label{eq:wavenumber}
\ee
and energies
\be
	E_n\equiv\frac{(\hbar k_n)^2}{2M}.
	\label{eq:boxenergies}
\ee

When we substitute the expansion coefficients $\psi_n$, eq.~(\ref{eq:kringel}), into the expression, eq.~(\ref{eq:psiofxandt}), for the wave function, we find the propagation equation
\be
	\psi(x,t)=\int\limits_0^L\drm y\,G_{\text{box}}(x,t|y,t=0)\,\varphi(y),
\ee
in which
\be
	G_{\text{box}}(x,t|y,t=0)\equiv\frac2L\sumlim_{n=1}^\infty\sin(k_nx)\sin(k_ny)\ehoch{-iE_nt/\hbar}
\ee
is the Green's function of the box.

The quadratic dispersion relation, eq.~(\ref{eq:boxenergies}), together with the definition, eq.~(\ref{eq:wavenumber}), of the wave number, yields phases
\be
	E_n\frac t\hbar=2\pi n^2\frac tT
	\label{eq:doppelkringel}
\ee
with characteristic time
\be
	T\equiv\frac{4ML^2}{\hbar\pi}.
	\label{eq:boxrevivaltime}
\ee
Hence, the Green's function of a particle in a box reads
\begin{equation}
	G_{\text{box}}(x,t|y,t=0)=\frac{2}{L}\sum\limits_{n=1}^{\infty} \sin(k_n x)\sin(k_n y)\exp\left(-2\pi i n^2\frac{t}{T}\right).
	\label{eq:starquest}
\end{equation}
In complete analogy to the Talbot effect, we find that for $t=T$ the phases, eq.~(\ref{eq:doppelkringel}), are integer multiples of $2\pi$. At this time the Green's function is identical to the one at time $t=0$. Consequently, the initial wave function revives, that is $\psi(x,t=T)=\psi(x,t=0)=\varphi(x)$. The similarity of the phases in the Talbot effect and the box problem justifies the use of the same symbol $T$ for the Talbot time and the revival time.

We conclude this section by noting that the quadratic dispersion relation, eq.~(\ref{eq:boxenergies}), is also the origin of quantum carpets \cite{bib:friesch, bib:physworld} observed in a box \cite{bib:berryklein,bib:berry} and in many other quantum systems \cite{bib:berry2, bib:marzoli}.

\subsection{Relation to the Talbot propagator}

It is instructive to cast the sum of products of sine functions in the propagator, eq.~(\ref{eq:starquest}), into a slightly different form. For this purpose we recall the relation
\begin{equation}
	\sin(k_n x)\sin(k_n y) = \frac{1}{4}\left[e^{i k_n(x-y)}+e^{-i k_n(x-y)}-e^{i k_n(x+y)}-e^{-i k_n(x+y)}\right].
\end{equation}
The terms with negative phase can be combined with the ones having positive phase by extending the summation in eq.~(\ref{eq:starquest}) to negative values, that is
\begin{equation}
	G_{\text{box}}(x,t|y,t=0)=\frac{1}{2L}\sum\limits_{n=-\infty}^{\infty} \left[e^{-i k_n(x-y)}-e^{-i k_n(x+y)}\right]\exp\left(-2\pi i n^2\frac{t}{T}\right).
	\label{eq:expexp}
\end{equation}
The term $n=0$ does not contribute since $k_0=0$.

When we introduce the Green's function
\begin{equation}
	G_{\text{T}}(x,t|y,t=0)\equiv\frac{1}{2L}\sum\limits_{n=-\infty}^{\infty}\exp\left[-in\frac\pi L(x-y)\right]\exp\left(-2\pi i n^2\frac{t}{T}\right),
	\label{eq:gbox}
\end{equation} 
eq.~(\ref{eq:expexp}) takes on the form
\begin{equation}
	G_{\text{box}}(x,t|y,t=0) = G_{\text{T}}(x,t|y,t=0)-G_{\text{T}}(x,t|-y,t=0).
\end{equation}
Hence, the propagator of the box is the difference of two Green's functions with starting points $y$ and $-y$, both leading to $x$ at time $t$. This difference ensures the boundary condition that the wave function vanishes at the walls.

When we compare the Green's function $G_{\text{T}}$, eq.~(\ref{eq:gbox}), to the Talbot propagator $G_{\text{Talbot}}$, eq.~(\ref{eq:gtalbot}), we find that they are identical provided $L=d/2$.

\section{Time evolution and autocorrelation function}
\label{sec:wavepackets}

So far we have concentrated on the time evolution of two specific quantum systems: a wave periodic in space and a particle confined to a box. Both systems have shown quadratic phase factors and complete revivals. In the present section we generalize this treatment and show that quantum systems with a discrete energy spectrum can display the same phenomena.

\subsection{Definition}

We consider a discrete superposition
\be
	\ket{\psi(t=0)}=\sumlim_{n=0}^\infty\psi_n\ket n
\ee
of energy eigenstates $\ket n$ with energy eigenvalues $E_n$. Possible quantum systems are the inter-nuclear motion of a diatomic molecule \cite{bib:vrakking}, a Rydberg electron \cite{bib:yeazell}, the center-of-mass motion of an atom in a standing light wave \cite{bib:raithel} or a single mode of an electromagnetic field in a cavity \cite{bib:eberly}.

Due to the time evolution the energy eigenstates accumulate phases $E_nt/\hbar$ leading to the state
\be
	\ket{\psi(t)} = \sumlim_{n=0}^\infty\psi_n\ehoch{-iE_nt/\hbar}\ket n.
\ee
The autocorrelation function
\be
	\left|S(t)\right|^2\equiv\left|\braket{\psi(t=0)}{\psi(t)}\right|^2
	\label{eq:autocorr}
\ee
with
\be
	S(t)\equiv\sumlim_{n=0}^\infty W_n\ehoch{-iE_nt/\hbar}
	\label{eq:soft}
\ee
is a measure for the overlap between the initial state $\ket{\psi(t=0)}$ and the state $\ket{\psi(t)}$ at time $t$. This overlap is a sum over all quantum numbers $n$ of weights $W_n\equiv|\psi_n|^2$ with phases $E_nt/\hbar$.

\subsection{Quadratic expansion of the energy spectrum}

When the occupation probability $W_n$ has a dominant maximum around a quantum number $\nbar$ and the energy spectrum is only slightly changing with $n$, we can expand $E_n$ into a Taylor series
\be
	E_n\approx E_{\nbar}+\left.\frac{\partial E_n}{\partial n}\right|_{n=\nbar}(n-\nbar)+\frac12\left.\frac{\partial^2E_n}{\partial n^2}\right|_{n=\nbar}(n-\nbar)^2,
\ee
retaining at most terms quadratic in $n-\nbar$. With the definitions
\be
	\left.\frac{\partial E_n}{\partial n}\right|_{n=\nbar}\equiv\hbar\frac{2\pi}{T_{\text{cl}}}
\ee
and
\be
	\frac12\left.\frac{\partial^2E_n}{\partial n^2}\right|_{n=\nbar}\equiv\hbar\frac{2\pi}{T},
\ee
where $T_{\text{cl}}$ and $T$ denote the classical period and the revival time, respectively, we arrive at the approximation
\be
	E_nt/\hbar\approx E_{\nbar}t/\hbar+\frac{2\pi t}{T_{\text{cl}}}(n-\nbar)+\frac{2\pi t}{T}(n-\nbar)^2
\ee
of the phases.
 
We substitute this expression into the definition, eq.~(\ref{eq:soft}), of $S(t)$, introduce the new summation index $m\equiv n-\nbar$ and arrive at
\be
	S(t)\approx\ehoch{-iE_{\nbar}t/\hbar}\,{\cal S}(t),
\ee
where
\be
	{\cal S}(t)\equiv\sumlim_{m=-\infty}^\infty\widetilde W_m\exp\left[-2\pi i\left(\frac{m}{T_{\text{cl}}}+\frac{m^2}{T}\right)t\right]
	\label{eq:cals}
\ee
with $\widetilde W_m=|\psi_{m+\nbar}|^2$. Here we have extended the lower bound $-\nbar$ of the summation to $-\infty$ since the dominant contributions arise for $m\approx0$.

In complete analogy to the Green's functions, eqs.~(\ref{eq:gtalbot}) and (\ref{eq:expexp}), of the Talbot effect or the particle in a box, the autocorrelation function, eq.~(\ref{eq:autocorr}), is determined by a sum where the summation index enters the phase in a quadratic way. However, in contrast to these two examples, we now also have a linear contribution providing two distinct time scales $T_{\text{cl}}$ and $T$.

\section{Fractional revivals}
\label{sec:fracrevivals}

In ref.~\cite{bib:leichtle1} we have devised a method to rewrite the sum ${\cal S}(t)$, eq.~(\ref{eq:cals}), in an exact way as to bring out the features of ${\cal S}(t)$ typical for the different time domains. We now concentrate on times
\be
	t=\ell\,T_{\text{cl}}+\Delta t=\frac qrT+\varepsilon_{q/r}T_{\text{cl}}+\Delta t
	\label{eq:ttepsilon}
\ee
that are close to a fraction $q/r$ of $T$ and are close to a large integer multiple $\ell$ of $T_{\text{cl}}$. The contribution $\varepsilon_{q/r}T_{\text{cl}}$ is a correction term, since in general we have
\be
	\ell\,T_{\text{cl}}\neq\frac qrT.
\ee
According to ref.~\cite{bib:leichtle1} we can cast ${\cal S}(t)$ into the form
\be
	{\cal S}(t=\frac qrT+\varepsilon_{q/r}T_{\text{cl}}+\Delta t)=\sumlim_{m=-\infty}^\infty {\cal W}_m^{(r)}{\cal I}_m^{(r)}(\Delta t)
	\label{eq:summewmim}
\ee
with the Gauss sums \cite{bib:lang}
\be
	{\cal W}_m^{(r)}\equiv\frac1r\sumlim_{p=0}^{r-1}\exp\left[-2\pi i\left(p^2\frac qr+p\frac mr\right)\right]
	\label{eq:wmrgausssumme}
\ee
and the shape functions
\be
	{\cal I}_m^{(r)}(\Delta t)\equiv\int\limits_{-\infty}^\infty\drm\mu\,\widetilde W(\mu)\exp\left\{-2\pi i\left[\left(\frac{\Delta t}{T_{\text{cl}}}-\frac mr\right)\mu+\left(\varepsilon_{q/r}+\frac{\Delta t}{T_{\text{cl}}}\right)\frac{T_{\text{cl}}}{T}\mu^2\right]\right\}.
	\label{eq:imrofdeltat}
\ee
Here $\widetilde W(\mu)$ denotes the continuous extension of $\widetilde W_m$.

The importance of this representation stands out most clearly for the example of a Gaussian weight function
\be
	\widetilde W(\mu)\equiv\sqrt{\frac{1}{2\pi\Delta n^2}}\exp\left[-\frac12\left(\frac{\mu}{\Delta n}\right)^2\right]
	\label{eq:gaussocc}
\ee
of width $\Delta n$.

In this case we can perform the integral, eq.~(\ref{eq:imrofdeltat}), and find the explicit expression
\be
	{\cal I}_m^{(r)}(\Delta t)=\widetilde{\cal N}(\Delta t)\exp\left[-\frac{\left(\Delta t-\frac mrT_{\text{cl}}\right)^2}{2\sigma_r^2(\Delta t)}\right]\exp\left[-i\frac{\left(\Delta t-\frac mrT_{\text{cl}}\right)^2}{2\sigma_i^2(\Delta t)}\right]
	\label{eq:starsharp}
\ee
for the shape function. Here we have introduced the complex amplitude
\be
	\widetilde{\cal N}(\Delta t)\equiv\frac1{\sqrt{1-i4\pi\Delta n^2(\varepsilon_{q/r}T_{\text{cl}}+\Delta t)/T}}
\ee
and the widths
\be
	\sigma_r^2(\Delta t)\equiv\left[\frac1{4\pi^2\Delta n^2}+4\Delta n^2\left(\frac{\varepsilon_{q/r}T_{\text{cl}}+\Delta t}{T}\right)^2\right]T_{\text{cl}}^2
	\label{eq:sigmar}
\ee
and
\be
	\sigma_i^2(\Delta t)\equiv\left[\frac1{16\pi^3\Delta n^2(\varepsilon_{q/r}T_{\text{cl}}+\Delta t)/T}+\frac1\pi\Delta n^2\frac{\varepsilon_{q/r}T_{\text{cl}}+\Delta t}{T}\right]T_{\text{cl}}^2
\ee
of the real and the imaginary Gaussians.

According to eq.~(\ref{eq:summewmim}) the sum ${\cal S}(t)$, determining the autocorrelation function, contains the product ${\cal W}_m^{(r)}{\cal I}_m^{(r)}(\Delta t)$ of the Gauss sum ${\cal W}_m^{(r)}$ and the shape function ${\cal I}_m^{(r)}(\Delta t)$. The latter consists of the product of a complex-valued square root, a real and an imaginary Gaussian. The Gaussians only take on non-vanishing values in the neighborhood of $\Delta t=mT_{\text{cl}}/r$. When the separation $T_{\text{cl}}/r$ between two neighboring Gaussians is larger than their width $\sigma_r$ they do not overlap. In this case the sum over $m$, that is, over the individual Gaussians, separates into a sequence of Gaussians.

Hence, the autocorrelation function in the neigborhood of a fractional revival, that is at a time $t=\frac qrT+\varepsilon_{q/r}T_{\text{cl}}+\Delta t$ consists of a sequence of Gaussians separated by $T_{\text{cl}}/r$ provided $r$ is odd or $2T_{\text{cl}}/r$ for $r$ even. This dependence on the odd-even-property of $r$ is a consequence \cite{bib:buch} of the Gauss sums ${\cal W}_m^{(r)}$, eq.~(\ref{eq:wmrgausssumme}). When the Gaussians do not overlap the $m$th term in the summation, eq.~(\ref{eq:summewmim}), represents the $m$th fractional revival.

When neighboring non-vanishing terms ${\cal I}_m^{(r)}(\Delta t)$ and ${\cal I}_{m'}^{(r)}(\Delta t)$ overlap, interferences between these terms arise. Then the phases of the complex Gaussian and the square root start to play an important role. Consequently, the sum ${\cal S}(t)$ exhibits a more complicated pattern.

\section{Factorization using wavepackets}
\label{sec:factorization}

In the preceding section we have cast the sum ${\cal S}(t)$ determining the autocorrelation function into a sequence of complex-valued Gaussians. We now show, that this form suggests a scheme to factorize numbers.

\subsection{General principle}
For this purpose we return to the non-overlap criterion and recall from eq.~(\ref{eq:sigmar}), that the width $\sigma_r$ of each Gaussian is different since it depends on time $\Delta t$. The minimal width
\be
	\sigma_r^{\text{(min)}}=\frac{T_{\text{cl}}}{2\pi\Delta n}.
\ee
occurs for the time $\Delta t_{\text{min}}$ defined by
\be
	\varepsilon_{q/r}T_{\text{cl}}+\Delta t_{\text{min}}=0.
\ee

If $\varepsilon_{q/r}=0$ the time $\Delta t_{\text{min}}$ of minimal width vanishes. Since the Gaussian with index $m=0$ also has its maximum at $\Delta t=0$, this Gaussian is of minimal width. Moreover, it has no overlap with neighboring ones provided $\sigma_r<\frac1rT_{\text{cl}}$. This condition puts the constraint
\be	
	\Delta n>\frac{r}{2\pi}
\ee
on the width $\Delta n$ of the Gaussian weight function $\widetilde W$, eq.~(\ref{eq:gaussocc}).

According to eq.~(\ref{eq:ttepsilon}), a vanishing correction term
\be
	\varepsilon_{q/r}=\ell-\frac qr\frac{T}{T_{\text{cl}}}=0
\ee
corresponds to the condition
\be
	\ell\cdot r=q\cdot N
\ee
where $N=T/T_{\text{cl}}$ is the number to be factorized.

Hence, a well-localized Gaussian at time $\ell\cdot T_{\text{cl}}$ indicates $\varepsilon_{q/r}=0$ and thus provides the factor $\ell$ of $N$.

We conclude this section by briefly discussing the case when $\ell$ is not a factor of $N$ and therefore $\varepsilon_{q/r}$ is non-zero. In this case the structure at $\Delta t=0$ is not a Gaussian of minimal width. Indeed, the widths of this Gaussian and its neighbors are so large, that they overlap considerably and the phase factors in $\widetilde N$ and in the imaginary Gaussian, eq.~(\ref{eq:starsharp}), lead to complicated interference structures. No clear fractional revival occurs at $\Delta t=0$.

\subsection{Simulation}
In order to test our predictions, we encode the number $N$ in the ratio $T/T_{\text{cl}}$ of the two time scales. We then search the autocorrelation function for maxima at integer multiples of $T_{\text{cl}}$.

\begin{figure}
\includegraphics[width=13cm]{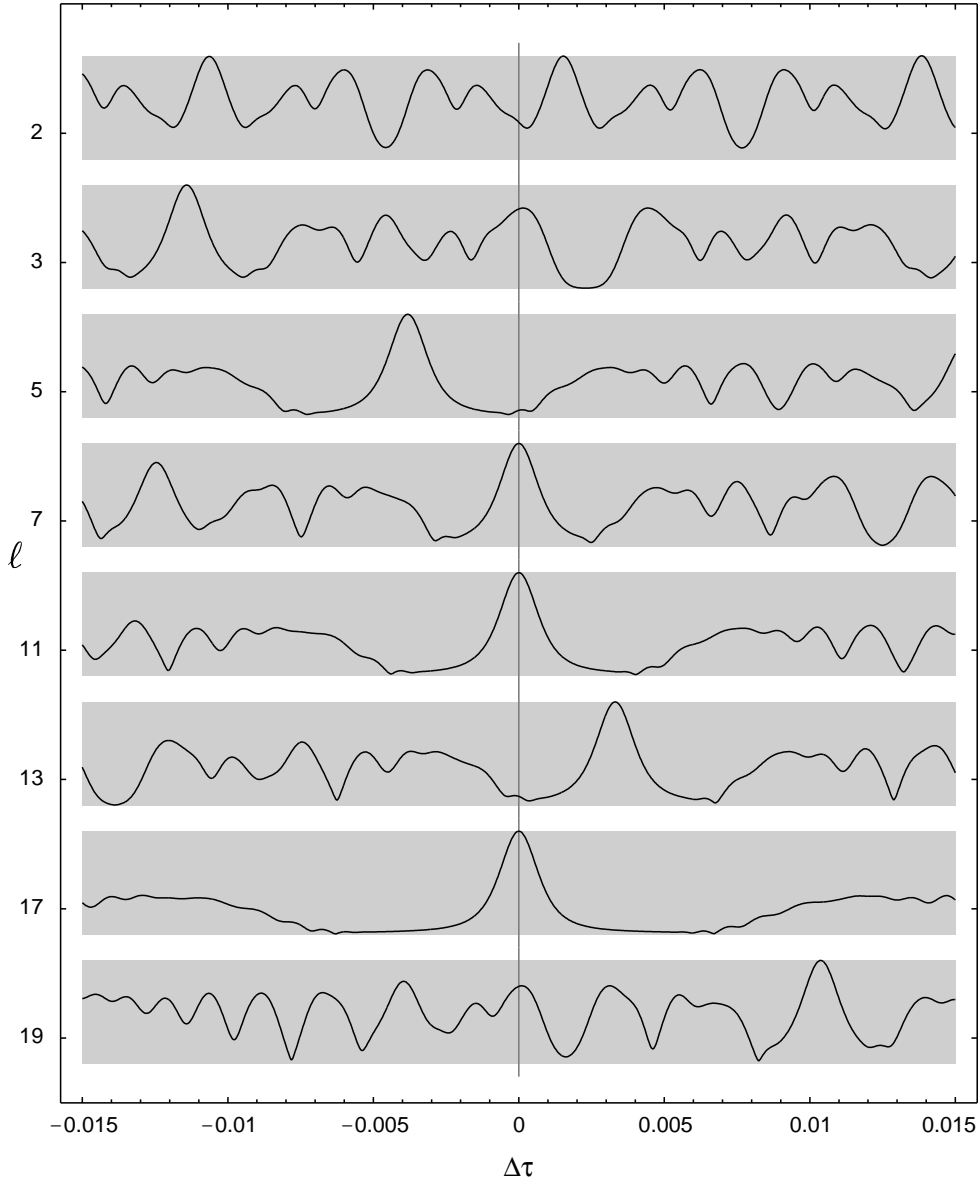}
\caption{Autocorrelation function $|{\cal S}_N(\tau=\ell+\Delta\tau)|^2$, defined in eq.~(\ref{eq:scaledauto}), as a function of dimensionless time $\Delta\tau$ in the vicinity of various integers $\ell=2,3,5\dots$. The goal is to find the factors of $N=1309=7\cdot 11\cdot 17$. The autocorrelation function has a maximum at an integer, that is at the origin of each horizontal axis, provided this integer is a factor of $N$. In the present case we clearly recognize $7$, $11$ and $17$ as factors.
\label{fig:revivals}}
\end{figure}

In fig.~\ref{fig:revivals} we show the autocorrelation function
\be
	\left|{\cal S}_N(\tau)\right|^2\equiv\left|\sumlim_{m=-\infty}^{\infty}W(m)\exp\left[-2\pi i\left(m+\frac{m^2}{N}\right)\tau\right]\right|^2
	\label{eq:scaledauto}
\ee
as a function of dimensionless time $\tau\equiv t/T_{\text{cl}}$ for the Gaussian weight function, eq.~(\ref{eq:gaussocc}). Here we have chosen the width $\Delta n=250$. The number we want to factorize is $N=1309=7\cdot 11\cdot 17$. Since the behavior of ${\cal S}_N(\tau)$ in the vicinity of an integer is important, we show $|{\cal S}_N(\tau)|^2$ around various integers. We recognize dominant maxima at $\tau=7$, $\tau=11$ and $\tau=17$, which are indeed the factors of $N$.

\section{Gauss sums and factorization}
\label{sec:gausssums}

The success of our factorization scheme relies on the interference of quadratic phase factors. Why not concentrate on the bare essentials of the method and eliminate the weight factors and the linear phase term altogether? We therefore consider the sum
\be
	s_N(n)\equiv\sumlim_{m=0}^{N-1}\exp\left(-2\pi im^2\frac nN\right)
	\label{eq:snn}
\ee
for fixed $N$ as a function of $n$.

This sum was also investigated in ref.~\cite{bib:curlicues} and gives rise to the so-called curlicues. The emphasis of ref.~\cite{bib:curlicues} was on the self-similarity of the emerging structures. However, in the present discussion we focus on the possibility of finding factors by considering the real and imaginary parts of this sum. 

\begin{figure}
\includegraphics[width=13cm]{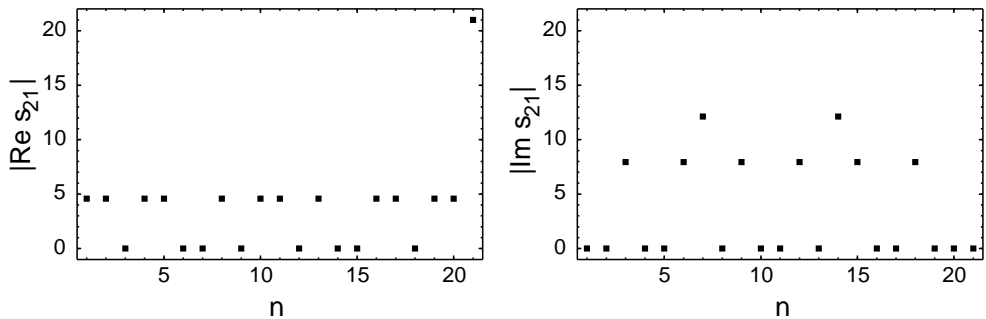}
\caption{Absolute value of real part (left) and imaginary part (right) of the Gauss sum $s_N(n)$, eq.~(\ref{eq:snn}), for $N=21=3\cdot 7$. The factors and their integer multiples appear in the imaginary part of $s_{21}(n)$. \label{fig:reimsum}}
\end{figure}

In fig.~\ref{fig:reimsum} we display real and imaginary parts of $s_{N=21}$. We recognize that the factors $3$ and $7$ appear as the periods in the imaginary part.

\section{Conclusions and outlook}
\label{sec:conclusions}

Interference of quadratic phase factors emerging in the Talbot effect, the particle in a box or in fractional revivals of wave packets has the potential to factorize large numbers. So far our technique relies solely on interference and does not make use of entanglement, a purely quantum mechanical degree of freedom. In a next step we want to combine the effect of the quadratic phases with the advantages entanglement can offer. We therefore have to consider composite quantum systems, such as highly dimensional spin systems, and develop generalized measurements (POVMs). Besides establishing a novel link to the problem of factorizing large numbers, this approach will povide new insight into the connections between quantum physics and number theory.

\acknowledgments
We thank I.Sh.~Averbukh, M.V.~Berry, H.~Maier and I.~Marzoli for many fruitful discussions. Moreover, two of us ({F.H.} and {W.P.S.}) are grateful to F.~De~Martini, P.~Mataloni and C.~Monroe for organizing a most stimulating summer school in the wonderful surroundings of Lake Como. We also thank the editors of these proceedings for patiently awaiting the completion of our manuscript. The work of {H.M.}, {F.S.S.}, {M.F.} and {W.P.S.} was supported by the Deutsche Forschungsgemeinschaft and by the European Commission through the IST network QUBITS.

\newcommand{\atque}{and }
\newcommand{\BY}[1]{#1,}
\newcommand{\REVIEW}[4]{{#1} \textbf{#2}, {#3} ({#4})}
\newcommand{\BOOK}[4]{\textit{#1} (#2, #3, #4)}

\end{document}